\documentclass[pra,showpacs,twocolumn]{revtex4}
\usepackage{amsmath,graphics,psfragx}

\newcommand{\eps}{\varepsilon}
\newcommand{\beq}{\begin{equation}}
\newcommand{\eeq}{\end{equation}}
\newcommand{\bea}{\begin{eqnarray}}
\newcommand{\eea}{\end{eqnarray}}

\newcommand{\mr}[1]{\ensuremath{\mathrm{#1}}}

\newcommand{\intd}{\ensuremath{\int\!\mathrm{d}}}

\newcommand{\intdl}[2]{\ensuremath{\int^{#2}_{#1}\!\mathrm{d}}}

\def\halb{\ensuremath{\frac{1}{2}\,}}

\def\fB{\ensuremath{f_\mr{B}}}

\def\T2b{\ensuremath{T_\mr{2B}}}

\renewcommand{\inf}{\infty}
\def\bfr{{\bf r}}

\def\r{{\bf r}}

\def\k{{\bf k}}

\def\x{{\bf x}}

\def\Lop{\ensuremath{\hat{\mathcal{L}}}}                 
\def\hGP{\hat h_\mr{\scriptscriptstyle GP}}              

\def\hpsidr{\hat \psi^\dagger (\bfr)}
\def\hpsi{\hat \psi (\bfr)}                   
\def\hpsid{\hat \psi^\dagger (\bfr)}

\def\halph{\hat \alpha}                                 
\def\halphd{\hat \alpha^\dagger}
\def\dpsi{\delta\hat\psi(\bfr)}                 
\def\dpsid{\delta\hat\psi^\dagger(\bfr)}

\def\psiip{\psi_i^{\scriptscriptstyle (\!+\!)}} 
\def\psiim{\psi_i^{\scriptscriptstyle (\!-\!)}} 

\def\nt{\ensuremath{\tilde{n}(\bfr)}}
\def\mt{\ensuremath{\widetilde{m}(\bfr)}}
\def\nth{\ensuremath{\tilde{n}}}

\def\eps{\epsilon}
\def\ceps{\varepsilon}
\def\cphi{\varphi}
\def\kB{\ensuremath{k_\mathrm{B}}}

\def\wtrap{\ensuremath{\omega_\bot}}

\def\Bec{Bose-Einstein condensation}

\def\BdG{Bogoliubov-de Gennes}

\def\GPE{Gross-Pitaevski\u{\i} equation}
\def\gGPE{generalized Gross-Pitaevski\u{\i} equation}

\def\cf{\mbox{c.\,f.\ }}
\def\ie{\mbox{i.\,e.\ }}

\begin{document}


\title{Coherence properties of the two-dimensional Bose-Einstein condensate}
\author{Christopher Gies}
\email{cluso@physics.otago.ac.nz}
\author{D.\ A.\ W.\ Hutchinson}
\email{hutch@physics.otago.ac.nz} \affiliation{Department of Physics, University of
Otago, P.O. Box 56, Dunedin, New Zealand}

\begin{abstract}
We present a detailed finite-temperature Hartree-Fock-Bogoliubov (HFB) treatment of the
two-dimensional trapped Bose gas. We highlight the numerical methods required to obtain
solutions to the HFB equations within the Popov approximation, the derivation of which we
outline. This method has previously been applied successfully to the three-dimensional
case and we focus on the unique features of the system which are due to its reduced
dimensionality. These can be found in the spectrum of low-lying excitations and in the
coherence properties. We calculate the Bragg response and the coherence length within the
condensate in analogy with experiments performed in the quasi-one-dimensional regime
[Richard \emph{et al.}, Phys.\ Rev.\ Lett.\ \textbf{91}, 010405 (2003)] and compare to
results calculated for the one-dimensional case. We then make predictions for the
experimental observation of the quasicondensate phase via Bragg spectroscopy in the
quasi-two-dimensional regime.
\end{abstract}

\pacs{03.75.Hh, 05.30.Jp, 67.40.Db}

\maketitle

\section{Introduction}

\Bec\ (BEC) in (quasi-) two-dimensional systems has only recently been obtained in the
laboratory \cite{Goerlitz2001_2D,Grimm2003_2dBEC}. Thus, many properties have yet to be
explored both experimentally and theoretically. We present an investigation of an
isotropic two-dimensional BEC with the aim of providing detailed predictions for
comparison with future experiments.

The manner in which dimensionality can fundamentally alter the physics of a system is
clearly apparent in the Mermin-Wagner-Hohenberg theorem, which forbids a spontaneously
broken symmetry with long range order in a homogeneous two-dimensional system
\cite{Mermin1966,Mermin1968,Hohenberg1967}. In terms of the coherence function
$G^{(1)}(\x,\x')$, this means that $\lim_{|\x-\x'|\rightarrow \inf} G^{(1)}(\x,\x') \neq
0$, which can be seen as the definition of BEC \cite{Penrose1956,Yang1962}, is impossible
for $T>0$ in a uniform two-dimensional system. Thus, in a two-dimensional Bose gas BEC
cannot occur at finite temperatures. Phase fluctuations make the formation of a globally
coherent phase impossible. Despite this, a different transition of the
Kosterlitz-Thouless (KT) type \cite{Kosterlitz1973,Kosterlitz1974,Berezinski1971} to a
state with an analytical decay in the coherence function is possible in the ideal system.

With confinement in a harmonic trap, the modified density of states allows the 2D system
to Bose condense. Nevertheless, below the critical temperature there is a large phase
fluctuating regime in which the superfluid is best described as a quasicondensate
\cite{Popov}. Unlike a true BEC, phase coherence only extends over regions of a size
smaller than the extent of the condensate, characterized by the coherence length. This
regime has been referred to as the KT phase, although the physical state of the
interacting system in this phase fluctuating regime has yet to be thoroughly
investigated. Phase fluctuations can enter the uniform gas in the form of
vortex/antivortex pairs, or topological charges, which unbind at the point of the KT
transition. Thus, the phase fluctuating state may well be a regular lattice of pairs of
opposite topological charges in the sense of the KT phase, but this is not the only
possibility and further investigation is required.

In a previous publication \cite{Cluso1} we have discussed how the semi-classical
approximation fails to describe BEC consistently in two dimensions and have shown results
to prove that these problems can be removed by applying the more complex
Hartree-Fock-Bogoliubov (HFB) formalism. The aim of the present publication is to present
a detailed and more complete discussion of the properties of a two-dimensional BEC as is
possible within the HFB-Popov approach. Our emphasis lies on the coherence properties
which are crucial for the question of whether the superfluid state is best described as a
BEC or as a quasicondensate. In the following section we outline the HFB formalism and
explain our methods of obtaining solutions. Then, in Section \ref{sec_results}, we
present our results, such as the density profile of the condensate and non-condensate,
the excitation spectrum and the coherence function. In Section \ref{ssec_coh} we present
the momentum profile and coherence length of a phase fluctuating condensate, indicating
how these could be measured in forthcoming experiments. Our work is concluded in Section
\ref{sec_conclusion}.

\section{Formalism}
\subsection{Mean-field theory and HFB-Popov equations}

The time-independent, second quantized form of the grand-canonical many-body Hamilton
operator for our system is given by
\begin{equation}\label{mf_hamil2}
 \begin{aligned}
    \hat H =& \intd^2 r\; \hpsid\, \left(\hat h(\bfr)-\mu \right)\, \hpsi \\
            & +\frac{g}{2}\,\intd^2 r\; \hpsid \hpsid\, \hpsi \hpsi~.
 \end{aligned}
\end{equation}
Here, $\hat h(\bfr)=-\frac{\hbar^2}{2m}\,\Delta+U_\mr{trap}(\bfr)$ is the single particle
Hamiltonian with the external potential  $U_\mr{trap}$ of the atom trap, and $g$ is the
coupling parameter that characterizes interparticle scattering. For collision processes,
we assume a hard sphere potential within the usual pseudo-potential approximation
\cite{HuangStatMech}, \ie $V(\bfr-\bfr') = g\,\delta^{(2)}(\bfr-\bfr')$. For a dilute gas
this is a good approximation, however care must be taken in determining the coupling
constant $g$. Usually it is derived from an approximation to the two-body T-matrix in the
zero-energy and zero-momentum limit, as appropriate for scattering processes in an
ultra-cold system. In three dimensions, the two-body T-matrix for a dilute gas is well
described within the s-wave approximation, $g = 4\pi \hbar^2 a_\mr{3D}/m$, where
$a_\mr{3D}$ is the s-wave scattering length. In two dimensions, however, the two-body
T-matrix vanishes at zero energy \cite{Lee2002a}. Therefore, many-body effects introduced
by the surrounding medium must be taken into account when studying two-dimensional gases.
For a trapped gas, this leads to a spatially dependent coupling parameter $g(\r)$.
Furthermore, the exact form of the coupling strength depends on the tightness of the
confinement in the axial direction. With the parameters from \cite{Goerlitz2001_2D},
using the terminology of \cite{Lee2002a}, we consider this system to be in the quasi-2D
regime. Therefore, for the calculations undertaken in this work, we use the following
approximation to the many-body T-matrix at zero temperature for the coupling parameter
\cite{Lee2002a}:
\begin{equation}\label{cp:g2d_spatial}
    g(\r) =  - \frac{4\,\pi \hbar}{m} \,
    \frac{1}{\ln\big(n_c(\r)\,g(\r)\, m a_\mr{2D}^2/4 \hbar^2\big)}~.
\end{equation}
The scattering length $a_\mr{2D}$ in the quasi-2D regime is given by $a_\mr{2D} =
4\,\sqrt{\pi/B} \:l_z\,e^{-\sqrt{\pi} \,l_z/a_{3D}}$, $B\approx 0.915$. This result was
first obtained by Petrov \emph{et al.}\ \cite{Petrov2001long,Petrov2000a} by considering
the 2D scattering problem. We will present a detailed study of interactions in the 2D
Bose condensed system elsewhere \cite{Cluso3}.

We decompose the Bose field operators, in the standard fashion
\cite{Griffin1996b,Hutchinson2000a}, into classical and fluctuation parts, $\hat
\psi(\bfr) \simeq \langle \hat\psi(\bfr) \rangle + \delta\hat\psi(\bfr) =
\Psi_0(\bfr)+\delta\hat\psi(\bfr)$, where the condensate wave function $\Psi_0(\r)$ is
normalized to the number of particles in the ground state, \ie $\intd^2r\:
|\Psi_0(\bfr)|^2 = N_0$. The Hamiltonian \eqref{mf_hamil2} can then be diagonalized by a
unitary transformation to the quasiparticle operators $\halph_i,\,\halphd_i$,
$\delta\hat\psi(\bfr) = \sum_i \big(\halph_i\, u_i(\bfr) - \halphd_i\, v_i^*(\bfr)\big)$,
yielding the HFB-Hamiltonian
\begin{equation}\label{mf_Hhfb_diag}
 \begin{split}
  \hat H_\mr{HFB} =& \intd^2r\; \Psi_0(\bfr)\,\left(\hat
  h(\bfr)-\mu+\halb g(\r)\, n_c(\bfr) \right)\,\Psi_0(\bfr)\\
   & + \sum_i E_i\, \halphd_i \halph_i - C
 \end{split}
\end{equation}
where $\Lop=\hat h(\bfr)-\mu+2\,g(\r)\, n(\bfr) \label{mf_L_op}$ and $n_c(\r)$, $\nt$ and
$n(\bfr)=n_c(\bfr)+\nt$ are the condensate, non-condensate and total densities,
respectively. The functions $u_i,\,v_i$ are referred to as quasiparticle amplitudes, and
$E_i$ are the quasiparticle energies. The first term in \eqref{mf_Hhfb_diag} is the
condensate part and merely a $c$-number. The second term is the Hamiltonian for
non-interacting quasiparticles and is formally equivalent to the case of the harmonic
oscillator. The constant energy shift $C$ arises from the Bogoliubov transformation
\cite{Griffin1996b} and from terms left over from the quartet operator averages of the
fluctuation operators which are factorized in a fashion analogous to Wick's theorem
\cite[{\S}4.2]{Morgan2000a}. However, this energy shift has no impact on the solution of the
HFB equations.

The form \eqref{mf_Hhfb_diag} of the Hamiltonian requires that the order parameter obeys
the \gGPE\ (GPE)
\begin{equation}\label{mf_fullgGPE}
    \Big( \hat h(\bfr)-\mu \Big)\, \Psi_0(\bfr) +
    g(\r)\,\big( n_c(\bfr)+2\,\nt\,
    \big)\,\Psi_0(\bfr)
     = 0
\end{equation}
and that the quasiparticle amplitudes and energies obey the coupled \BdG\ (BdG) equations
\begin{equation}\label{mf_BdG_full}
 \begin{split}
  \Lop\,u_i(\bfr) -g(\r)\, \Psi_0(\bfr)^2 \, v_i(\bfr) &= E_i\, u_i(\bfr)\\
  \Lop\,v_i(\bfr) -g(\r)\, \Psi^*_0(\bfr)^2 \, u_i(\bfr)
   &= - E_i\, v_i(\bfr)~,
 \end{split}
\end{equation}
so as to eliminate off-diagonal terms in the quasiparticle field operators. The BdG
equations determine the elementary excitation modes of the condensate. We refer to
\eqref{mf_fullgGPE}, together with \eqref{mf_BdG_full}, as the HFB equations. Note that
we have taken the Popov approximation by neglecting the anomalous average of the
fluctuation operator, $\mt = \langle \dpsi \dpsi \rangle$, whereby avoiding divergence
problems of this quantity and the occurrence of a gap in the excitations spectrum
\cite{Griffin1996b,Hutchinson2000a}. Once the BdG equations are solved, the
non-condensate density $\tilde n = \langle \dpsid \dpsi \rangle$ can be obtained by
populating the quasiparticle states,
\begin{equation}\label{mf_nth-uv}
 \nt = \sum_i
    \fB(E_i)\,\big(|u_i(\bfr)|^2+|v_i(\bfr)|^2 \big)+|v_i(\bfr)|^2~,
\end{equation}
where the quasiparticle distribution function with the inverse temperature $\beta$ is
given by
\begin{equation}\label{mf_qp_distrfn}
    \fB(E_i) = \langle \halphd_i \halph_i \rangle =
                \frac{1}{z^{-1} e^{\beta E_i} -1}~.
\end{equation}
Here, the fugacity $z$ is determined by the difference between the chemical potential
$\mu$ and the condensate eigenvalue $\lambda$, $z = e^{\beta (\mu-\lambda)}$, since the
quasiparticle energies are measured relative to the condensate \cite{Morgan2000a}. To a
good approximation, we can use the result for the non-interacting gas, \ie
\begin{equation}\label{mf_fugacity}
    z^{-1} = 1+\frac{1}{N_0}~.
\end{equation}
The system we consider has a finite number of atoms. The fugacity fulfills the practical
purpose of preventing the number of thermal atoms from exceeding the total atom number
and, hence, the condensate density from becoming negative in our numerical calculations.

In order to study coherence properties, we calculate the normalized first order
correlation, or coherence function, which can be written in terms of the field operators
as \cite{Glauber_corfkt}
\begin{equation}\label{ph_g1}
 g^{(1)}(\bfr,\bfr') = \frac{\langle\hat\psi^\dagger(\bfr)
\hat\psi(\bfr')\rangle} {\sqrt{\langle\hat\psi^\dagger(\bfr) \hat\psi(\bfr)\rangle
\langle\hat\psi^\dagger(\bfr') \hat\psi(\bfr')\rangle}}~.
\end{equation}
Given the decomposition of the field operator, the coherence function can be expressed in
terms of the off-diagonal densities
\begin{align}
     n_c(\r,\r') &= \Psi_0^*(\r)\Psi_0(\r')\\
    \nth(\r,\r') &= \langle \delta\hat\psi^\dagger(\r) \delta\hat\psi(\r')\rangle~.
\end{align}
The latter can be calculated from the off-diagonal version of \eqref{mf_nth-uv}. Using
the above for the correlation function, \eqref{ph_g1} gives
\begin{equation}\label{nm_g1}
    g^{(1)}(\r,\r') = \frac{n_c(\r,\r') + \nth(\r,\r')}
                          {\sqrt{n(\r)\,n(\r')} } ~.
\end{equation}
The correlation function is related to the momentum spectrum of the condensate by a
simple Fourier transformation, \ie
\begin{equation}\label{ph_nk}
    n(\k) = \langle \hat \phi^\dagger(\k) \hat \phi(\k) \rangle =
    \intd^2r \, \mr{d}^2r'\; e^{i \k \cdot(\r-\r')}\,
               \langle \hpsidr \hat \psi(\r') \rangle~,
\end{equation}
with $\hat \phi(\k)$ and $\hat \phi^\dagger(\k)$ being the field operators in momentum
space. This implies that coherence properties can be directly measured in an experiment,
as has been done in \cite{Richard2003a} for the quasi-one-dimensional case by means of
Bragg spectroscopy. In a Bragg experiment, the propagation speed of the light field is
determined by the detuning of the crossed laser beams \cite{Blakie2002a}. Therefore, the
spectral response of the condensate is measured as a function of the detuning. To
establish the relationship with the momentum distribution, we use the relation between
the detuning $\delta$ and the momentum within the condensate plane $p_\bot$ for a
$n$-photon process,
\begin{equation}\label{ph_momentumdetune}
    \delta = \frac{n\, k_L p_\bot}{2\pi\, m} ~,
\end{equation}
where $k_L=2\pi/\lambda$, $\lambda$ is the wavelength of the light field ($780.02\,$nm
for Rubidium \cite{Richard2003a}, $589\,$nm for Sodium \cite{Wilkinson1996a}), and $m$
the mass of the atoms.

%
%
%
%
\subsection{Numerical Methods}\label{ssec_num}

We discuss some aspects important to the solution of the finite temperature HFB
equations. The trapping frequency in the axial direction is sufficiently large so that
the dynamics in this dimension are frozen out ($\hbar\omega_z > \kB T$). In the radial
plane, we consider an isotropic trapping potential with the radial frequency $\wtrap$,
$U_\mr{trap} = m \wtrap^2 r^2 /2$. Thus, our system is cylindrically symmetric and we can
effectively treat the problem as one-dimensional upon changing to cylindrical
coordinates. We scale all equations to computational units, \ie lengths by the oscillator
length, $a_0 = \sqrt{\hbar/m \wtrap}$, and energies by the Rydberg of energy, $E_0 =
\hbar \wtrap/2$.

The calculation follows a self-consistent, iterative scheme, as proposed in
\cite{Griffin1996b}. First, the GPE is solved with the non-condensate density set to
zero. Taking this calculated condensate density, the BdG equations are solved to obtain
the quasiparticle modes. These are then populated through the quasiparticle distribution
function \eqref{mf_qp_distrfn}, with the sum of all the excited particles yielding the
thermal density. With the non-condensate density now known, we go back and solve the
generalized GPE and the whole process is repeated until convergence.

To begin, we expand the order parameter in a set of basis states. A convenient basis for
this problem is given by the eigenstates of the 2D harmonic oscillator, since the
single-particle Hamiltonian is diagonal in this basis. To take the cylindrical symmetry
into account, we write the eigenfunctions of the oscillator problem $\hat
h_\mr{osc}=-\Delta+r^2$ in terms of the Laguerre polynomials $L_n^m$,
\begin{equation}\label{nm_2dosc_chi}
    \chi_{n,m}(r,\cphi) = \frac{1}{\sqrt{\pi\: \Gamma(1+m)\,\binom{n+m}{n}}}\;
                          r^m e^{-\frac{r^2}{2}+i m \cphi}\, L_n^m(r^2)~,
\end{equation}
with the eigenenergies $E_{n,m}=2(2n+m+1)$. The quantum number $m$ defines the angular
momentum. Since the condensate ground state has zero angular momentum, for the solution
of the GPE merely the $m=0$ subspace must be considered. In order to numerically solve
the GPE, we use an optimization routine with a Thomas-Fermi profile as the initial guess.

The solution of the BdG equations follows the method described in \cite{Hutchinson2000a}.
In a first step, the BdG equations \eqref{mf_BdG_full} are decoupled by a transformation
to the auxiliary functions $\psi_i^{\scriptscriptstyle (\!\pm\!)}(r) = u_i(r) \pm
v_i(r)$. Omitting spatial dependencies, this leads to
\begin{equation}\label{nm_BdG_psi}
  \begin{aligned}
    \big(\hGP-\mu\big)^2 \psiip + 2g n_c \big(\hGP-\mu\big) \,\psiip &= E_i^2\,
    \psiip\\
    \big(\hGP-\mu\big)^2 \psiim + 2g \big(\hGP-\mu \big) n_c\,\psiim &= E_i^2\,
    \psiim~,
  \end{aligned}
\end{equation}
where $\hGP \equiv \hat h + g\,(n_c + 2\nth)$ is the Hamiltonian in the generalized GPE
\eqref{mf_fullgGPE}. The auxiliary functions $\psi_i^{\scriptscriptstyle (\!\pm\!)}(r)$
are then expanded in the basis set in which $\hGP$ is diagonal. This basis we term the
Hartree-Fock (HF) basis and it is obtained by the full solution of the generalized GPE,
\ie
\begin{equation}\label{nm_hf_gpe}
\Big( \hGP - \mu \Big) \, \phi_\alpha(r) = \ceps_\alpha \, \phi_\alpha(r)~,
\end{equation}
where $\{\phi_\alpha(r)\}$ is the HF basis. The primary advantage of this basis is that
all excited states are by definition orthogonal to the condensate, after the lowest
momentum state, which is the condensate state itself, has been removed from the basis
set. Both the calculation of the HF basis, as well as the solution of the decoupled BdG
equations are only linear problems, since the condensate density is given from the
solution of the GPE, and can be solved in a straightforward manner. The eigenvalue
problem corresponding to the BdG equations is block-diagonal with no overlap between the
subspaces of different angular momentum, so that the solution to \eqref{nm_BdG_psi} can
be obtained separately in each subspace. The thermal density then follows from
\eqref{mf_nth-uv} by summing up the contributions from all angular momentum subspaces.

Naturally, the number of basis states used in the discrete, quantum mechanical
calculation is limited by an upper energy cutoff, $\eps_\mr{cut}$, which must be
introduced consistently in all angular momentum subspaces. To account for the
contributions above the energy cutoff, we use the semi-classical approximation
\cite{Reidl1999a,Mullin1998a}, so that
\begin{equation}\label{nth_num}
 \begin{split}
    \nth(r) = \sum_{i} \nth_i^\mr{qm}(r) \times \Theta(\eps_\mr{cut}-E_i) +
    \intdl{\eps_\mr{cut}}{\inf}E\;
    \nth^\mr{sc}(E,r)~.
 \end{split}
\end{equation}
The contribution $\nth^\mr{qm}_i(r)$ below the cutoff is given by the addent in
\eqref{mf_nth-uv}, 
and above the cutoff by the semi-classical equation, with the Heaviside function
$\Theta$, and again omitting spatial dependencies,
\begin{equation}\label{mf_nth_sc}
 \begin{split}
    \tilde n^\mr{sc} &= \frac{m}{2\pi\hbar^2} \left\{
     \fB(E)+\halb - \frac{E}{2\, \sqrt{E^2+(g n_c)^2}}\right. \\
     & \quad \left.\times \,\Theta\left( E-\sqrt{\left(U_\mr{trap}-\mu+2g n\right)^2 -
        \left(g n_c\,\right)^2}\, \right) \right\}~.
 \end{split}
\end{equation}

\section{Results}\label{sec_results}

We will now present the results of our numerical calculation. We consider a sample of
2000 sodium atoms that are trapped in a harmonic potential with the parameters of the
experiment by G\"orlitz \emph{et al.}\ \cite{Goerlitz2001_2D}. The radial trapping
frequency is $\wtrap = 2\pi\times 790\,$Hz. Unless otherwise stated, all quantities are
expressed in dimensionless form.

\subsection{Thermal density and condensate population}

\subsubsection{Density profiles}
\begin{figure} 
\center
\includegraphicx[overwritepfx=true,width=.6 \columnwidth]{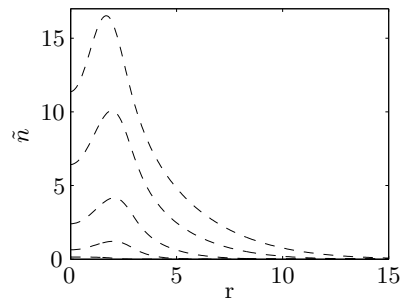}
\caption{Non-condensate density at $T/T_c = $ 0, 0.1, 0.25, 0.5 and 0.75 (from bottom to
top). The lowest line corresponds to the quantum depletion.}\label{fig:nths}
\end{figure}

\begin{figure}  
\center \rule{0 pt}{3mm}
\includegraphicx[overwritepfx=true,width=4.2cm]{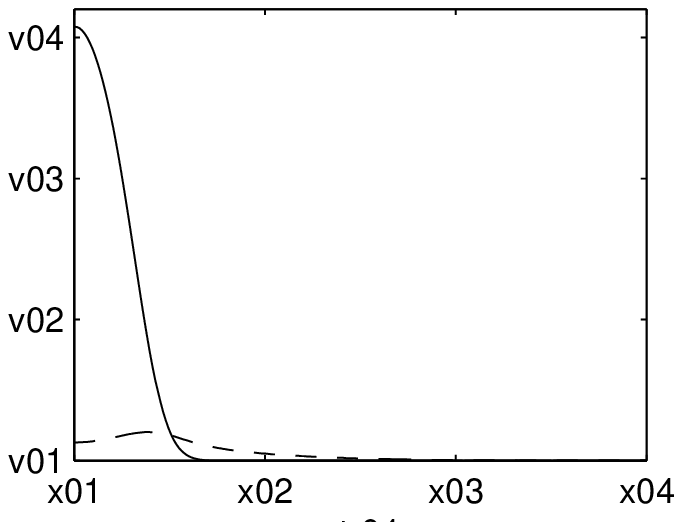}
\hspace{-4.2mm}
\includegraphicx[overwritepfx=true,width=4.2cm]{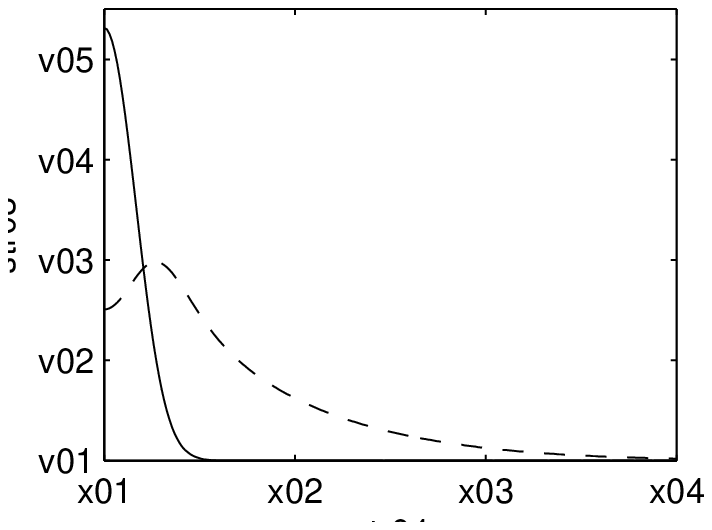}
\caption{Condensate (solid) and non-condensate density (dashed) at 0.5 and
0.9$\,T_c$.}\label{fig:nth-nc}
\end{figure}

Figure \ref{fig:nths} shows the thermal density at different temperatures. The
temperature dependent term in \eqref{mf_nth-uv} leads to the formation of the
characteristic off-center peak of the non-condensate density. It is located at the edge
of the condensate due to the repulsion of the thermal atoms by the condensate. This is
depicted in Figure \ref{fig:nth-nc}, where the two densities $n_c$ and $\tilde n$ are
plotted together. In comparison to the rapidly decaying condensate density, the thermal
density has a long tail. Thus, the condensate is relatively dense with a sharp peak
within the diffuse thermal cloud. The tail of the thermal cloud becomes longer as the
temperature increases, while the condensate radius does not change significantly even if
the number of condensate atoms drops by an order of magnitude. Note that the long tail
contains a large number of atoms despite its low density because the spatial integral is
weighted by a factor of $r$ ($r^2$ in three dimensions). The slight change in the size of
the condensate can also be seen in the shift of the non-condensate peak towards the trap
center with increasing temperature, remaining located at the edge of the condensate. Note
that even at zero temperature there is a small fraction of excited atoms due to the
temperature-independent quantum depletion term in \eqref{mf_nth-uv}.

From the existence of a well defined condensate and non-condensate density we can already
infer coherence information. Following the argument in \cite{Petrov2000a,Cluso1}, a
measure of phase fluctuations is given by $\langle \hat \delta^2 \rangle \approx \tilde
n/n_c$, where $\hat \delta$ is the phase fluctuation operator in the alternative
decomposition $\hat\psi(r) \simeq \sqrt{\hat n(r)}\: e^{i \hat \delta(r)}$. Phase
fluctuations become important when $\langle \hat \delta^2 \rangle \gtrsim 1$. Thus, as
long as $n_c > \tilde n$, these fluctuations are suppressed in the system.

%

\subsubsection{Ground state population}

\begin{figure}[h]  
\center
\includegraphicx[overwritepfx=true,width=.7 \columnwidth]{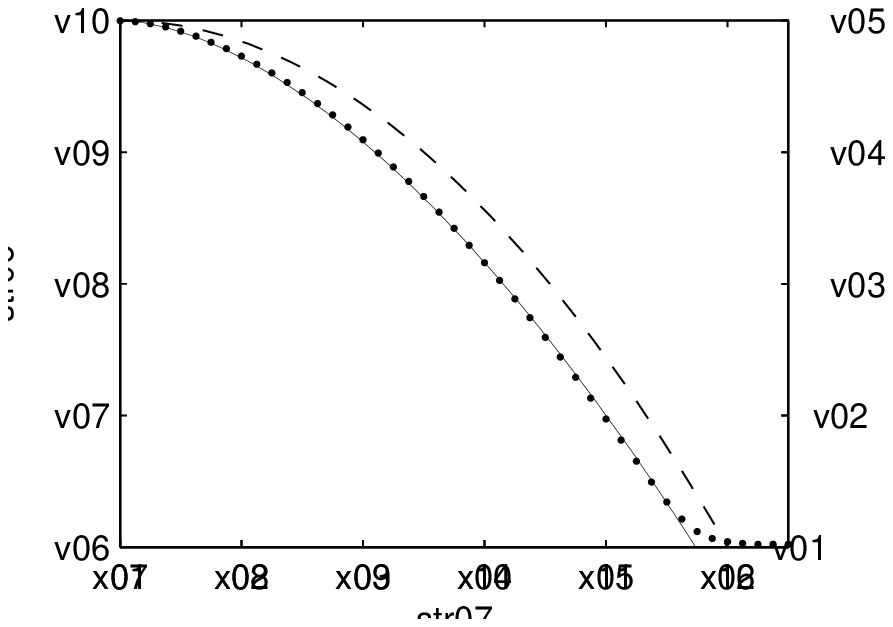}
\caption{Condensate population versus temperature. The solid line corresponds to a fit to
\eqref{res_population}, the dashed line shows the ideal gas power law dependence. The
points are results from the HFB-Popov calculation.}\label{fig:population}
\end{figure}

In Figure \ref{fig:population} the condensate population is shown as a function of
temperature. The results are compared to the case of the trapped ideal gas, where the
population is determined by a power law expression. We fit the following functional form
to the numerical data:
\begin{equation}\label{res_population}
    \frac{N_0}{N} = 1 - \left( \frac{T}{\bar{T}_c} \right)^\beta,\quad
    \text{where}\quad \bar{T}_c = \alpha  T_c~.
\end{equation}
In the case of the ideal gas, $\beta = 2$ and $\alpha = 1$. In the fit the critical
temperature is reduced by a factor of about 5\% with $\alpha \approx 0.95$. This shift
has two contributions: The finite size of the system reduces the critical temperature
\cite{Giorgini1996a}, but it is also modified by the interactions. This second
contribution as been extensively discussed in the literature, see the recent publication
\cite{Boris_sevenloops} and references therein. For the exponent we find $\beta \approx
1.70$, which is 15\% smaller than for the ideal trapped gas. With these values,
\eqref{res_population} parameterizes our data very well except near the critical
temperature, where finite size effects are significant and the exact method of
determining the shift of the chemical potential from the condensate eigenvalue becomes
important.
In the ideal gas the chemical potential is zero at the point of the phase transition, and
therefore the transition point is strictly defined. In the interacting gas, the chemical
potential depends implicitly on the non-condensate \cite{Hutchinson2000a} and, when this
becomes large, the transition point becomes smeared out. Technically, the occurrence of
the finite temperature tail can be explained through the fugacity term in the
quasiparticle distribution function \eqref{mf_qp_distrfn}, which is explicitly given by
\eqref{mf_fugacity}. The term $\propto 1/N_0$ prevents the condensate population from
becoming negative. However, this expression for the fugacity is only approximate and the
shape of the tail and the speed with which it approaches zero depends on the explicit
choice of the fugacity term at temperatures around $\bar{T}_c$.


\subsection{Condensate excitations}

The low-lying collective or elementary excitation modes of the condensate, determined by
the solution of the BdG equations, are of interest because they reflect certain
fundamental symmetry properties of the system, as well as being easily accessible to
experiments.

\begin{figure}  
\center
\includegraphicx[overwritepfx=true,width=.99 \columnwidth]{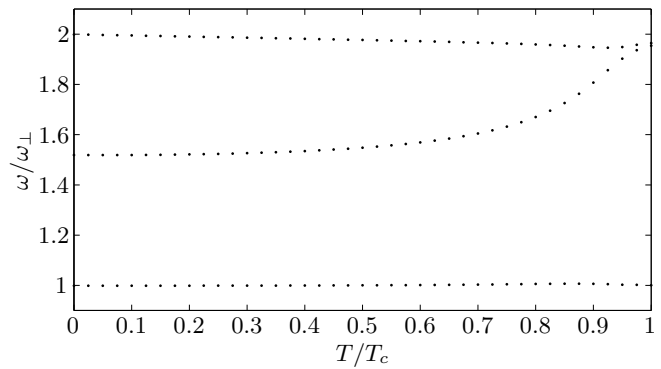}
\caption{Low-lying excitation modes of the condensate as a function of temperature. The
uppermost line corresponds to the \emph{breathing mode} with angular momentum quantum
number $m=0$, the middle line to the \emph{quadrupole mode} with $m=2$. The lowest line
is the \emph{Kohn mode}, $m=1$, which lies constantly at the trapping
frequency.}\label{fig:excitations}
\end{figure}

Figure \ref{fig:excitations} shows the modes with angular momentum $m=0,1,2$ as a
function of temperature. Each of the three branches corresponds to the lowest
quasiparticle energy eigenvalue in the lowest three, separated, angular momentum
subspaces, in which the BdG equations are solved, \cf Section \ref{ssec_num}.
\paragraph{Breathing mode.}
The breathing mode corresponds to an oscillation of the condensate radius and lies at a
frequency that is twice the trapping frequency. As shown in \cite{Pitaevskii1997a}, this
is due to a hidden symmetry of the many-body Hamiltonian with a
$\delta^{(2)}$-interaction potential and a harmonic trapping potential in two dimensions.
As the temperature increases, the non-condensate density grows and starts to constitute a
deviation from the harmonic oscillator potential in the effective potential of the \GPE,
so that the frequency shifts slightly from $2\omega_\bot$. The effective potential is
weakened by the presence of the static thermal atoms so that the frequency decreases. If
the dynamics of the thermal cloud were included in the calculation \cite{Morgan2003a},
then the full symmetry of the Hamiltonian would be restored and the mode frequency would
remain precisely at $2\hbar\wtrap$.

\paragraph{Kohn mode.} The Kohn mode corresponds to a center of mass oscillation
 of the whole condensate.
Less effected by the perturbation to the harmonic potential of the static thermal cloud,
it remains very constant at the trapping frequency, as is predicted by the generalized
Kohn theorem \cite{Dobson1994}. However, looking closer at Figure \ref{fig:excitations},
one may see a slight increase in the frequency near the critical temperature as the
effective potential becomes less harmonic and, therefore, breaks the Kohn theorem. In our
calculation we treat the thermal cloud as stationary. An inclusion of the full dynamics
of the thermal cloud would, again, ensure the Kohn mode remains constant at all
temperatures \cite{Proukakis1998a,Hutchinson2000a,Morgan2003a}.

\paragraph{Quadrupole mode.} The quadrupole mode is the only low-lying mode which depends
strongly upon the temperature. The frequency of this mode could thus, in principle, be
used as a measure of the temperature of the 2D gas.

With increasing temperature, all three frequencies smoothly approach the frequencies of
the non-interacting gas and the breathing and quadrupole mode become degenerate. Using a
local density approximation with the relation $\mu=g[n_c(r)+2\tilde n(r)]$ for the
uniform gas in the Hartree-Fock approximation \cite[{\S}8.3]{PethickBook}, it is easy to
show that the BdG equations for $n_c(r) = 0$, given by
\begin{equation}\label{res:BdG_nc0}
    \Big( \hat h(r) - \mu + 2g\tilde n(r) -E_i \Big)\,u_i(r)= 0~,
\end{equation}
recover the energies of the harmonic oscillator. In the case that there is still a
condensate, the total density in the region where $n_c(r) \neq 0$ is approximately
constant just below the critical temperature, so that the mean-field energy only
constitutes a near constant shift to the trapping potential and, hence, only slightly
alters the eigenfrequencies of the trap.

We would briefly like to draw comparison with the three-dimensional case where the
frequency spectrum looks very similar \cite{Hutchinson2000a,Hutchinson1998a}. The
striking difference is the breathing mode which is temperature dependent in three
dimensions, whereas it is a feature of the two-dimensional system to have breathing
oscillations with a universal energy of $2\hbar\wtrap$.

\subsection{Coherence properties}\label{ssec_coh}

\subsubsection{Correlation function}

\paragraph{Interacting gas.}

\begin{figure}  
\center
\includegraphicx[overwritepfx=true,width=.7 \columnwidth]{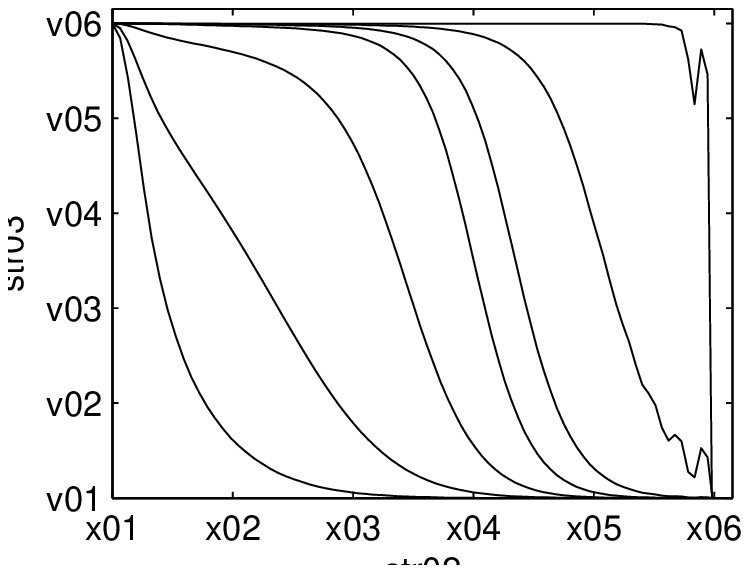}
\caption{Correlation function $g^{(1)}$ for the non-interacting Bose gas at different
temperatures $T/T_c$: 0, 0.025, 0.05, 0.1, 0.5, 0.9 and 1 from right to
left.}\label{fig:g1_popov}
\end{figure}

In Figure \ref{fig:g1_popov} the correlation function $g^{(1)}(0,r)$ is depicted at
various temperatures. The $r$-axis has been scaled by the size of the condensate. This is
not the Thomas-Fermi radius, but we choose a minimal allowed condensate density in such a
way that the whole condensate at zero temperature is phase coherent.

The decay of the correlation function allows for a characterization of the gaseous
system. At low temperatures the correlation function has a constant value throughout the
extent of the condensate, indicating a truly coherent Bose-Einstein condensed phase with
off-diagonal long-range order. Algebraic decay is associated with the KT phase and, at
intermediate temperatures, the superfluid must be identified as a quasicondensate. At
very high temperatures, clearly visible for the highest temperature in Figure
\ref{fig:g1_popov}, the coherence function decays exponentially, showing that long-range
order is lost completely.

At very low temperatures the correlation functions show some unphysical oscillations that
are purely numerical noise. At low temperatures $n(r) \approx n_c(r)$. In the limit
$\tilde n\equiv 0$ the correlation function \eqref{ph_g1} is given by the Heaviside
function $\Theta(1-r/r_\mr{con})$. However, there is a small contribution from the
quantum depletion of the condensate that smoothes the sharp corner as $g^{(1)}$ drops to
zero, causing a loss of numerical accuracy as we divide two very similar small numbers in
\eqref{nm_g1}.

The coherence length can be extracted by measuring the full width at half maximum (fwhm)
of $g^{(1)}$ and is shown in Figure \ref{fig:lphi_bragg}.

%

\paragraph{Non-interacting gas.}

The off-diagonal density matrix is known in closed analytical form for the
non-interacting gas. In can be determined by means of the inverse Laplace transform of
the zero-temperature Bloch density matrix \cite{Brandon_g1,Brandon_some}. Its explicit
form in two dimensions at temperature $T$ is given by
\begin{multline}\label{res_g1noninteract}
    g^{(1)}(\r,\r',T) = \sum_{j=1}^\inf \frac{e^{j\mu/T}}{\pi(1-e^{-2j/T})}
    \times\\
     \exp\left( -\frac{|\r+\r'|^2}{4} \tanh(j/2T)
     -\frac{|\r-\r'|^2}{4} \coth(j/2T)\right)~.
\end{multline}
We find the chemical potential $\mu=\mu(T)$ for the trapped non-interacting gas by
solving $\sum_{n=0}^{\inf} \fB(E_n=n+1,\mu,T)\,(n+1) - N = 0$ with respect to $\mu$.
Here, $\fB=\left[(1+N_0^{-1} ) e^{\beta (E_n-\mu)} -1\right]^{-1}$ is the Bose-Einstein
distribution function \emph{with the fugacity factor} \eqref{mf_fugacity} that takes into
account the number of condensate particles from the HFB calculation. Without this
fugacity factor, finite size effects, taken into account in the HFB calculation, would be
neglected and, therefore, the comparison would be between two approaches based on
different assumptions. The expression \eqref{res_g1noninteract} for the correlation
function is exact at all temperatures. Numerically, the infinite sum can be calculated up
to any required accuracy.

We compared our code against these exact results for the non-interacting gas. At all
temperatures the HFB results agree perfectly with the correlation function calculated
from \eqref{res_g1noninteract}, implying that the numerics works well even at high
temperatures. Deviations would indicate an insufficiency in the basis set or inaccuracy
due to an insufficient fineness or range of the computational grid.

\begin{figure}  
\center
\includegraphicx[overwritepfx=true,width=.9 \columnwidth]{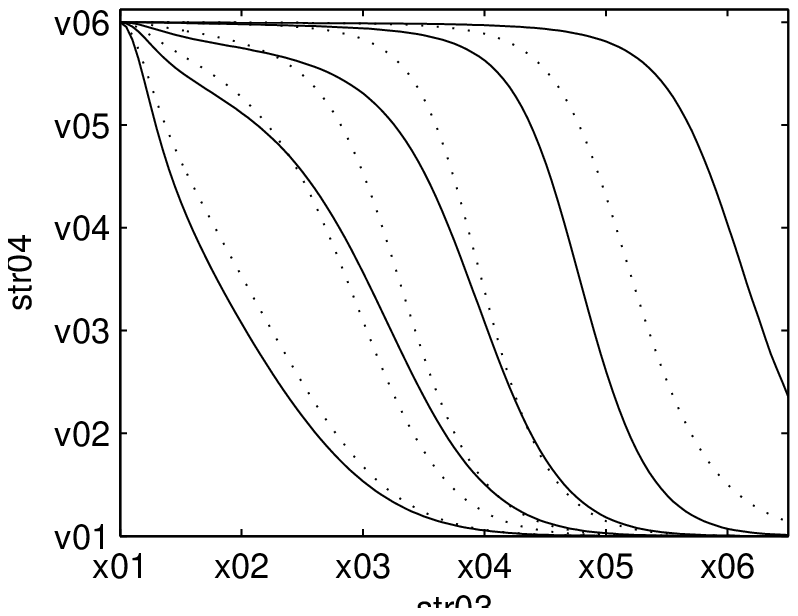}
\caption[Comparison of the correlation function for the interacting and non-interacting
gas]{Correlation function $g^{(1)}$ for the non-interacting Bose gas at different
temperatures $T/T_c$: 0.025, 0.1, 0.5, 0.8 and 0.95 from right to left. The solid lines
correspond to the HFB-Popov results for the interacting gas, the dotted lines represent
the exact result for the non-interacting gas
\eqref{res_g1noninteract}.}\label{fig:cp_g1ni}
\end{figure}

The influence of the interactions can be seen in Figure \ref{fig:cp_g1ni}. The solid
lines show the coherence function for the interacting gas, compared to the exact
non-interacting gas equation, shown as dotted lines. We see that interactions increase
the coherence length in a large part of the temperature regime. At $0.8\,T_c$ and above,
however, the coherence length of the interacting gas is decreased relative to the
non-interacting gas. This can be explained by the effect of the mean-field interaction on
the condensate radius. At high temperatures ($N_c$ small) the radius of the interacting
condensate is approximately the same as the radius of the non-interacting condensate,
given by the size of the lowest harmonic oscillator state. At low temperatures, however,
the condensate population is large and mean-field effects broaden the condensate.
Correspondingly, the coherence function of the interacting condensate is broader. If the
spatial coordinate was scaled by the size of the condensate as in Figure
\ref{fig:g1_popov}, the plot would show that interactions always reduce the range of
coherence.

%

\subsubsection{Momentum spectrum}

\begin{figure}
\includegraphicx[overwritepfx=true,width=5cm]{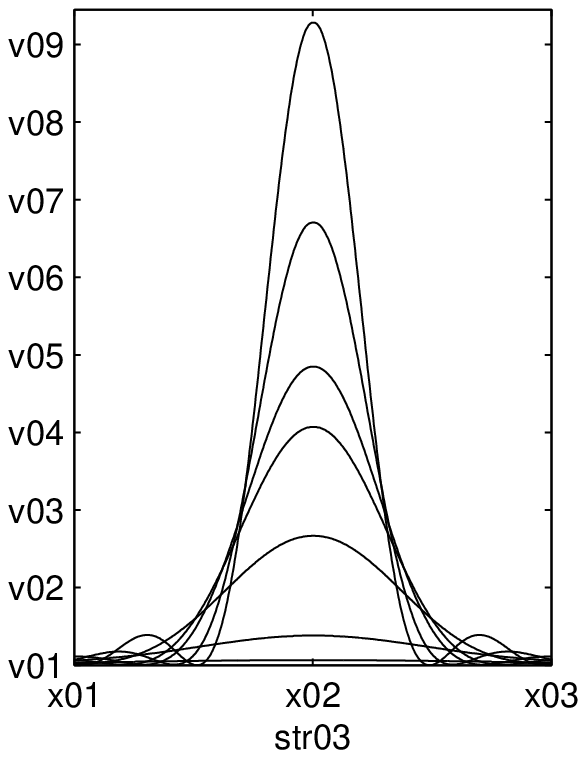}
\caption{Momentum spectrum, with intensity corresponding to the fraction of scattered
 atoms as function of the Bragg detuning $\delta$. Temperatures $T/T_c$ from top: 0, 0.025, 0.05,
 0.1, 0.5, 0.9 and 1.}\label{fig:momentumspectrum}
\end{figure}

Figure \ref{fig:momentumspectrum} shows the momentum spectrum corresponding to
\eqref{ph_nk}, as it could be measured by means of Bragg spectroscopy. It has been
calculated by Fourier transforming the correlation function shown in Figure
\ref{fig:g1_popov}. On the ordinate is the detuning of the Bragg laser beams, which is
directly proportional to the momentum of the atoms, \cf \eqref{ph_momentumdetune}. The
highest peak corresponds to the lowest temperature where the momentum distribution of the
atoms is narrowest. With increasing temperature the spectrum is broadened. An
experimental setup is limited by its resolution at low temperatures, because of the
decrease of the spectral width.

\begin{figure}  
\center
\includegraphicx[overwritepfx=true,width=.7 \columnwidth]{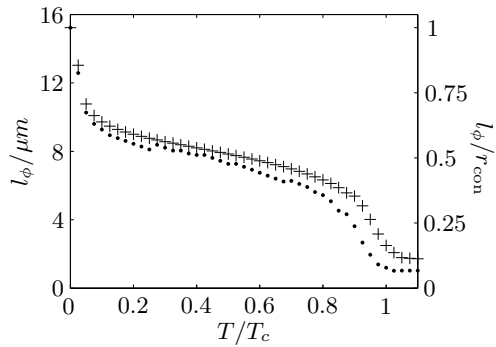}\rule[0cm]{5mm}{0mm}
\caption{Coherence length of the condensate, shown in both real units and scaled by the
extension of the condensate. The upper line ({\scriptsize +}) has been calculated from
the momentum spectrum (left panel), the lower line ($\cdot$) is the data from Figure
\ref{fig:g1_popov}.}\label{fig:lphi_bragg}
\end{figure}

In Figure \ref{fig:lphi_bragg}, the coherence length obtained from the momentum spectra
in Figure \ref{fig:momentumspectrum} is plotted. We determine the coherence length by
fitting the momentum profile and measuring the half width of the fit. A Gaussian provides
a good fit at temperatures ${\scriptstyle\lesssim\,} 0.9\,T_c$. Above a small crossover
regime, the momentum profiles at temperatures higher than $0.925\,T_c$ fit more closely
to a Lotentzian. Thus, the data points in Figure \ref{fig:lphi_bragg} correspond to the
fwhm of a Gaussian or a Lorentzian fit, depending on which gives better agreement with
the data.

In the same graph the lengths are compared to those obtained from the results shown in
Figure \ref{fig:g1_popov}. Qualitatively the results agree with each other, although
those obtained from $g^{(1)}$ lead to somewhat smaller values for the coherence length.
Also one can see that the extracted half widths of the correlation function are subject
to a slight inaccuracy, whereas the widths calculated from the fits to the momentum
profile result in a smooth line over the whole temperature regime. Note that the
coherence length is to some extent a matter of personal definition, as e.$\,$g.\ we could
have chosen $1/e$ rather than the fwhm.

Looking at the decay of the coherence length in Figure \ref{fig:lphi_bragg}, we can
distinguish three different regimes. Close to zero temperature, the slope is very steep
and the coherence length decreases to a third of the condensate size by about $0.1\,T_c$.
Then, up to about $0.8\,Tc$, $l_\phi$ decreases monotonically, but much more slowly. From
there up to the critical temperature, the coherence length again drops rapidly.

A decreasing coherence length directly implies a loss in the global phase coherence of
the superfluid phase. A true Bose-Einstein condensate cannot be said to exist when the
phase of the order parameter fluctuates on a length scale significantly smaller than the
extent of the condensate. At this point we should instead refer to a quasicondensate.
From Figure \ref{fig:g1_popov} we see that the coherence length drops smoothly.
Therefore, it is difficult to determine an exact point on the temperature scale where the
transition from a true condensate to a quasicondensate takes place. At about $0.5\,T_c$
the coherence length has dropped to about half of the maximal value. The maximal value we
can use to determine the spatial extent of the condensate, indicated on the right axis in
Figure \ref{fig:lphi_bragg}. The treatment in \cite{Petrov2000a} predicts a value of
approximately $0.4\,T_c$ for our parameters for phase fluctuations to become dominant.
Our result is, therefore, consistent with \cite{Petrov2000a}, although the coherent phase
seems to persist at slightly higher temperatures.

\subsubsection{Comparison to 1D}

Similar behaviour has been observed in calculations for the one-dimensional Bose gas at
finite temperatures \cite{Ghosh2004a}. Looking at the coherence function presented by
Ghosh, we see that in the one-dimensional case the coherence length drops even more
rapidly than in the two-dimensional case, showing that phase fluctuations become much
more dominant as the dimension is reduced further. In 1D, the temperature range between
0.3 and 0.5$\,T_c$ has about the same coherence properties as the range around $0.9\,T_c$
in our 2D calculation. Ghosh identifies the 1D phase at temperatures as low as $0.1\,T_c$
as a quasicondensate with large phase fluctuations. From an examination of Figure
\ref{fig:g1_popov}, we see that, at this temperature in the 2D case, even if the
coherence length has decreased slightly, there is still a large proportion of the
condensate where $g^{(1)}$ is constantly 1, indicating that the system is essentially a
phase coherent BEC.

In 1D the Lorentzian momentum profile has been found to be characteristic of the
phase-fluctuating quasicondensate \cite{Gerbier2003a} and has been used as an the
identifying signature of such a phase \cite{Richard2003a}. However, we are convinced that
the shape change we observe is not a signature of a phase fluctuating condensate, but the
effect of the fugacity term as $N_0$ goes to zero. Furthermore, from looking at the
correlation function in Figure \ref{fig:g1_popov}, we would expect the phase fluctuations
to become important, indicating the presence of a quasicondensate, at about $0.5\,T_c$,
long before the momentum profile becomes Lorentzian in character.\\

\section{Conclusion}\label{sec_conclusion}

We have used the HFB formalism to investigate the finite-tem\-pera\-ture physics of a
Bose-Einstein condensate confined to a two-dimensional geometry. Unlike the three
dimensional case, phase fluctuations must be taken into consideration at comparatively
lower temperatures. In a regime below the critical temperature they destroy the global
coherence of the condensate and the superfluid state is best described as a
quasicondensate. In the HFB formalism phase fluctuations are included via the
contribution to the non-condensate density from low-energy quasiparticles. We have shown
that the formalism is not only applicable in the strictly phase coherent regime, but also
that the quantities obtained, such as the single-particle off-diagonal density matrix,
allow for a quantitative analysis even in the phase fluctuating regime. Our work is
consistent with \cite{Petrov2000a}, although we find that, within the HFB treatment, the
pure condensate phase persists to higher temperatures.


The coherence length of the condensate can be determined from its correlation function or
the momentum profile. Following Aspect \emph{et al.}\ for the one-dimensional case
\cite{Aspect2003icols,Richard2003a}, we have calculated the Bragg spectrum for a
condensate in two dimensions. We found the values extracted for the coherence length to
be in qualitative agreement with those calculated for the one-dimensional Bose gas,
although a true BEC with global phase coherence still exists at much higher temperatures
than in the 1D case. The Bragg spectrum provides a clear signature of the quasicondensate
phase and we anticipate experimental efforts in this area in the near future.

\begin{acknowledgments}

The authors would like to acknowledge financial support from the Marsden and ISAT
Linhages Funds of the Royal Society of New Zealand, as well as a University of Otago
Research Grant. We thank Sam Morgan, Mark Lee and Brandon van Zyl for many useful
conversations during various exchange visits and subsequently.

\end{acknowledgments}


\end{document}